\documentclass[twocolumn]{revtex4}
\usepackage{graphicx}
\usepackage{dcolumn}
\usepackage{amsmath}

\begin{document}

\title{Nambu-Goldstone bosons characterized by the order 
parameter in spontaneous symmetry breaking
}

\author{Takashi Yanagisawa}

\affiliation{Electronics and Photonics Research
Institute, National Institute of Advanced Industrial Science and Technology (AIST),
Tsukuba Central 2, 1-1-1 Umezono, Tsukuba 305-8568, Japan
}


\begin{abstract}
We present explicitly a relation between the Nambu-Goldstone boson and the 
order parameter in non-relativistic systems
with spontaneous symmetry breaking.
We show that the Nambu-Goldstone bosons are characterized by
transformation property of the order parameter under symmetry
transformation of a system.
We give an explicit formula for the Nambu-Goldstone boson for a
general Lie group $G$, and then
the number of the Nambu-Goldstone boson is derived straightforwardly
from the form of the order parameter (the type of symmetry breaking).
We show that the Ward-Takahashi identity is modified in the presence of the
Nambu-Goldstone boson, where
the generalized Ward-Takahashi identity includes the coupling (the vertex 
function) between fermions and Nambu-Goldstone bosons.
The closed equation for the Green's functions of Nambu-Goldstone bosons 
is derived by introducing the fermion-Nambu-Goldstone boson vertex
function. 
Examples are given for $G=SU(2)$ (ferromagnetic), $U(1)$
(superconductor) and $SU(3)$ symmetry breaking.
\end{abstract}


\maketitle

\section{Introduction}

Symmetry is important for a better understanding of the laws of nature.
When the Lagrangian or the Hamiltonian is invariant under a symmetry 
transformation,
we have a conserved current and a conserved quantity.
When the Lagrangian is not invariant under some transformation, the
corresponding conservation of the current is violated.
There is often the case where the Lagrangian is invariant under a
symmetry transformation, but the state is not invariant under
this transformation.  This means that an asymmetric state is realized
in a symmetrical system.  This is called the spontaneous symmetry
breaking because it is a spontaneous process.
When a continuous symmetry is broken spontaneously, a massless boson,
called the Nambu-Goldstone boson (NG boson) 
emerges\cite{gold61,nam60,gold62}.
Two general proofs of their existence were then given in 
Ref.\cite{gold62,wein}.
The spontaneous symmetry breaking has been an interesting subject
since then in field 
theory\cite{col85,nam61,hig64,wei72,gol66,bra10,nie76,wat11,wat12,hid13,oda13,
wat14,nit15b}
and in the study of magnetism and superconductivity before 
then\cite{and84,g-l,bcs,abs88,whi06,lep74}.

The Ward-Takahashi identity follows from the invariance of the
Lagrangian\cite{war50,tak57}.
When the current conservation is violated by symmetry breaking,
the Ward-Takahashi identity is never followed.
The Ward-Takahashi identity is restored to hold, however, by means of
the existence of the Nambu-Goldstone boson.  This was examined
in Ref.\cite{nam61,gri94}.
The  Ward-Takahashi identity is modified
when the continuous symmetry is spontaneously broken.

Recently, the spontaneous symmetry breaking was classified into
two groups Type I and Type II\cite{nie76,wat11,wat12,hid13},
and the dispersion relation of the Nambu-Goldstone boson was
clarified following this classification.
However, the relation between the Nambu-Goldstone
boson and the order parameter is not clear since the theory
is primarily based on the algebra of conserved quantities.
The order parameter is important in the second-order phase
transition which is realized as a spontaneous symmetry breaking.
In this paper, we focus on the second-order phase transition,
and show that the Nambu-Goldstone bosons are fully characterized by the
transformation property of the order parameter $\Delta$
under symmetry trasformation of a system.

In this paper, we investigate the system with
an invariance under the 
continuous transformation group $G$ (compact Lie group). 
We focus on non-relativistic models in this paper.
The Nambu-Goldstone boson is expressed by means
of the bases of Lie algebra of $G$  once the order parameter $\Delta$
is expressed as the expectation value of a boson field or a product
of fermion fields.
A new proof is given to show that the Nambu-Goldstone boson indeed represents
a massless particle. 
Several proofs were given to show the existence of the Nambu-Goldstone
boson when a continuous symmetry is spontaneously broken\cite{wein}.
These proofs are, however, formal and abstract.  It is helpful to
give an explicit proof of the existence of the NG boson, and formulate
the NG boson by means of fermion or boson fields explicitly.

We introduce a small symmetry breaking term in the Lagrangian
(or the Hamiltonian) like the Zeeman term in a ferromagnet.
When the ground states are degenerate continuously, 
operators $Q_a$, generators of transformation, are not
well-defined in the Hilbert space.
The symmetry breaking term, namely, the external field is
introduced so that the ground state is unique and the matrix
elements of $Q_a$ are defined.
Lastly we take the vanishing limit of external field.

We also examine the Ward-Takahashi identity which is violated when
there is a spontaneous symmetry breaking.  
The Ward-Takahashi identity is restored by including a
contribution of the Nambu-Goldstone boson.  In other words,
the breaking of the Ward-Takahashi identity is compensated by
the inclusion of the Nambu-Goldstone boson. 


This paper is organized as follows.
In the next section, we give a formulation of spontaneous symmetry
breaking and give a formula for the Nambu-Goldstone boson $\pi_a$.
We first examine a fermion system.
We show that $\pi_a$ represents a massless boson.
We give several examples of spontaneous symmetry breaking.
In the section III, the Ward-Takahashi identity with the correction
from the Nambu-Goldstone bosons is investigated, where the NG 
boson-fermion coupling (vertex function) is introduced.
The equation for the Green's function of NG bosons is obtained
by using the NG boson-fermion vertex function.
We give a summary in the last section.

\section{Nambu-Goldstone boson}

\subsection{Invariant Lagrangians}
We consider models that are invariant under a continuous symmetry 
transformation of a Lie group $G$.
A fermion Lagrangian is given in the form,
\begin{equation}
\mathcal{L}_F= \psi^{\dag}i\Gamma^{\mu}\partial_{\mu}\psi+V(\psi),
\end{equation}
where $\psi$ represents a fermion field.
We can also examine a boson Lagrangian given as
\begin{equation}
\mathcal{L}_B=  \phi^{\dag}\left( i\hbar\frac{\partial}{\partial t}
+\xi(\nabla)\right)\phi-V(\phi),
\end{equation}
or the Lagrangian
\begin{equation}
\mathcal{L}_B = \frac{1}{2}\partial_{\mu}\phi^{\dag}\partial^{\mu}\phi-V(\phi),
\end{equation}
where $\phi$ is a scalar field,  $V(\phi)$ is the potential term and 
$\xi(\nabla)$ is the
dispersion relation.


We investigate a fermion system in the following.
When the Lagrangian is invariant under the transformation
$\psi\rightarrow \psi+\delta\psi$, we have the conserved current
\begin{equation}
j^{\mu}= \frac{\delta\mathcal{L}}{\delta(\partial_{\mu}\psi)}\delta\psi,
\end{equation}
with $\partial_{\mu}j^{\mu}=0$.
Let us denote the conserved currents as $j^{\mu}_a$ when there are several
conserved currents and the corresponding 
conserved
quantities as $Q_a$.
Let us consider a Lie group (transformation group) $G$ and corresponding
representation of fermion field $\psi$.
Let $g$ be the Lie algebra of the Lie group $G$.
We denote the basis set of the Lie algebra $g$ as $\{T_a\}$.
We assume that $T_a$ is hermitian.
The field transformation
$\psi\rightarrow \psi+\delta\psi$ is given by
\begin{equation}
\psi\rightarrow e^{-i\theta T_a}\psi =
\psi-i\theta T_a\psi+O(\theta^2),
\end{equation}
where $\theta$ is an infinitesimal parameter.
We write
\begin{equation}
\delta\psi = -i\theta T_a\psi,
\end{equation}
and define the conserved quantities as
\begin{equation}
Q_a = \int d{\bf r}J_a^0({\bf r}),
\end{equation}
where we set
\begin{equation}
J_a^{\mu} = \frac{1}{\theta}j_a^{\mu}.
\end{equation}
We put $\Gamma^0=1$ for simplicity to obtain
\begin{equation}
[Q_a,\psi]= -T_a\psi,
\label{qpsi}
\end{equation}
and
\begin{equation}
Q_a= \int d{\bf r} \psi^{\dag}T_a\psi.
\end{equation}

\subsection{Spontaneous Symmetry Breaking}

Let us introduce the term to the Lagrangian, which breaks the symmetry:
\begin{equation}
\mathcal{L}_{SB}= \lambda \psi^{\dag}M\psi,
\end{equation}
where $M$ is a c-number hermitian matrix in $\{T_a\}$. 
$\lambda$ is an infinitesimal real number and we let $\lambda\rightarrow 0$
at the end of calculations.
$\mathcal{L}_{SB}$ is the external field such the Zeeman term
in a ferromagnet.
We denote the total Hamiltonian including the symmetry 
breaking term as $H_T$.
We assume that the ground state of $H_T$ is unique, so that
we avoid the difficulty stemming from the degeneracy of
ground states\cite{ryd96}.
If the ground state is not unique, we must add another
symmetry breaking term to the Lagrangian to lift the degeneracy.

We define the order parameter $\Delta$ as the expectation value of this term:
\begin{equation}
\Delta = \langle \psi^{\dag}M\psi \rangle.
\end{equation}
We define that the symmetry generated by $Q_a$ with $[T_a,M]\neq 0$ 
is spontaneously
broken when $\Delta$ is finite $(\neq 0)$ in the limit $\lambda\rightarrow 0$.
The susceptibility $\chi_{\Delta}$ is define as
\begin{equation}
\chi_{\Delta} = \lim_{\lambda\rightarrow 0}\frac{\Delta}{\lambda}.
\end{equation}
$\chi_{\Delta}$ diverges when there is a spontaneous
symmetry breaking.

Under the transformation $\psi\rightarrow \psi-i\theta T_a\psi$, 
$\mathcal{L}_{SB}$ transforms to $\mathcal{L}_{SB}+\delta\mathcal{L}_{SB}$
where
\begin{equation}
\delta\mathcal{L}_{SB}= i\theta\lambda\psi^{\dag}[T_a,M]\psi.
\end{equation}
In this case, the current $j^{\mu}$ is not conserved:
\begin{equation}
\partial_{\mu}j_a^{\mu}= \delta\mathcal{L}_{SB}.
\end{equation}
Then we have
\begin{equation}
\partial_{\mu}J_a^{\mu}= i\lambda\psi^{\dag}[T_a,M]\psi.
\end{equation}
The divergence $\partial_{\mu}J^{\mu}_a$ is nothing but a Nambu-Goldstone
boson.  We define the Nambu-Goldstone boson as
\begin{equation}
\pi_a = i\psi^{\dag}[T_a,M]\psi.
\end{equation}
This means
\begin{equation}
\partial_{\mu}J_a^{\mu}= \lambda\pi_a.
\end{equation}
We show that $\pi_a$ indeed indicates a massless boson in the 
subsection 2.4.

Similarly, the Nambu-Goldstone in a boson system emerges.
We introduce the symmetry breaking term $\mathcal{L}_{SB}$ to the
Lagrangian $\mathcal{L}_B$.  For example, we add
\begin{equation}
\mathcal{L}_{SB} = \lambda(\phi+\phi^{\dag}).
\end{equation}
Examples of scalar field theories are discussed in the 
subsection 2.3.

\subsection{Examples of Symmetry Breaking}

We show several examples of spontaneous symmetry breaking on the basis of our
formulation in this subsection.
\subsubsection{Ferromagnetic transition}
We consider the fermion Lagrangian in Eq.(1) where
$\psi$ is given by a doublet of fermions:
\begin{eqnarray}
\psi(x)&=&\left(
\begin{array}{c}
\psi_{\uparrow}(x) \\
\psi_{\downarrow}(x) \\
\end{array}
\right).
\end{eqnarray}
Here $\psi_{\sigma}$ represents the annihilation operator of fermion
with spin $\sigma$.
The symmetry group is $G=SU(2)$ and the bases $\{T_a\}$ are given by
Pauli matrices $T_a=\sigma_a$ ($a=1,2$ and 3).
The structure constants are $f_{abc}=2\epsilon_{abc}$.
The Lagrangian is invariant under the transformations
\begin{equation}
\psi\rightarrow e^{-i\theta\sigma_a}\psi,
\end{equation}
for $a=1,2$ and 3.
The symmetry breaking term is given by the magnetization of
electrons:
\begin{equation}
\mathcal{L}_{SB} = \lambda\psi^{\dag}\sigma_3\psi
= \lambda(\psi^{\dag}_{\uparrow}\psi_{\uparrow}-
\psi^{\dag}_{\downarrow}\psi_{\downarrow}).
\end{equation}
When $\Delta\equiv \langle\psi^{\dag}\sigma_3\psi\rangle\neq 0$
in the limit $\lambda\rightarrow 0$, the symmetry is broken
spontaneously.
Since $[\sigma_1,\sigma_3]\neq 0$ and $[\sigma_2,\sigma_3]\neq 0$,
this term breaks the symmetry $\psi\rightarrow e^{-i\theta\sigma_a}\psi$
for $a=1,2$.
The Nambu-Goldstone bosons are
\begin{eqnarray}
\pi_1&=& i\psi^{\dag}[\sigma_1,\sigma_3]\psi= 2\psi^{\dag}\sigma_2\psi,\nonumber\\
\pi_2&=& i\psi^{\dag}[\sigma_2,\sigma_3]\psi=-2\psi^{\dag}\sigma_1\psi.
\end{eqnarray}
The excitation mode represented by $\pi_1$ and $\pi_2$ is spin-flip process,
that is, the spin-wave excitation.
We make a linear combination of $\pi_1$ and $\pi_2$ as
$\pi \equiv (i\pi_1-\pi_2)/4= \psi^{\dag}_{\uparrow}\psi_{\downarrow}$ and 
$\pi^{\dag} = (-i\pi_1-\pi_2)/4= \psi^{\dag}_{\downarrow}\psi_{\uparrow}$. 
Actually, there is only one Nambu-Goldstone boson $\pi$ in a ferromagnetic state.

\subsubsection{Antiferromagnetic transition}
In the case of antiferromagnetic transition, we divide the space into
two sublattices called A and B.
We adopt that electrons are on a bipartite lattice.
We denote the fermion fields on A and B sublattices as $\psi_A$ and
$\psi_B$, respectively.
We have $SU(2)$ symmetry in each sublattice.
The symmetry breaking term is
\begin{equation}
\mathcal{L}_{SB}= \lambda\psi^{\dag}_A\sigma_3\psi_A
+\lambda\psi^{\dag}_B\sigma_3\psi_B.
\end{equation}
The order parameters are $\Delta_A=\langle\psi^{\dag}_A\sigma_3\psi_A\rangle$
and $\Delta_B=\langle\psi^{\dag}_B\sigma_3\psi_B\rangle$ with the
constraint $\Delta_A+\Delta_B=0$ in the antiferromagnetic case.
In a similar way as in the ferromagnetic case, $\pi=(i\pi_1-\pi_2)/4$
(and its conjugate $\pi^{\dag}$) is the Nambu-Goldstone
boson in each sublattice. 
Thus we have two NG bosons $\pi_A$ and $\pi_B$ in this case.

\subsubsection{Scalar field theories}
(a) Single-component scalar field\\
Let us consider a complex scalar field model with the Lagrangian,
\begin{equation}
\mathcal{L}= \phi^{\dag}\left( i\hbar\frac{\partial}{\partial t}
+\frac{\hbar^2}{2m}\nabla^2+\mu\right)\phi(x)
-\frac{g_{\phi}}{2}\phi^{\dag}(x)\phi^{\dag}(x)\phi(x)\phi(x),
\end{equation}
where $g_{\phi}$ is the coupling constant, $x=(t,{\bf r})$ and 
we set $\hbar=1$.
The Lagrangian is invariant under the transformation
\begin{equation}
\phi\rightarrow e^{-i\theta}\phi.
\end{equation}
The conserved current is given by 
$j^0=i\phi^{\dag}\delta\phi= \theta\phi^{\dag}\phi$, and
$j^k=-(1/2m)(\partial_k\phi^{\dag})\delta\phi=
i\theta(1/2m)(\partial_k\phi^{\dag})\phi$ ($k=1$, 2 and 3).
We define $J^{\mu}=j^{\mu}/\theta$ so that
\begin{equation}
J^0= \phi^{\dag}\phi,~~ J^k= i\frac{1}{2m}\left(\partial_k\phi^{\dag}\right)
\phi.
\end{equation}
We include the symmetry breaking term:
\begin{equation}
\mathcal{L}_{SB}= \lambda (\phi+\phi^{\dag}).
\end{equation}
The divergence of the current is 
$\partial_{\mu}j^{\mu}=\delta\mathcal{L}_{SB}=-i\lambda\theta(\phi-\phi^{\dag})$.
The order parameter is
\begin{equation}
\Delta= \langle \phi+\phi^{\dag}\rangle.
\end{equation}
Since $\partial_{\mu}J^{\mu}=\lambda\pi$, the Nambu-Goldstone boson is
given as
\begin{equation}
\pi= i(\phi^{\dag}-\phi).
\end{equation}
\\
(b) Multi-component real scalar field\\
A symmetry breaking in a system with a multi-component model is
similarly examined.  For example, let us turn to a real three-component
scalar field $\phi= ^t(\phi_1,\phi_2,\phi_3)$ with the Lagrangian
\begin{equation}
\mathcal{L} = \frac{1}{2}(\partial_{\mu}\phi_i)(\partial^{\mu}\phi_i)
-V(\phi),
\end{equation}
where the summation convention is applied and $V(\phi)$ is
the potential.
We assume that the Lagrangian is invariant under the action of $G=SO(3)$.
The bases of the Lie algebra of $SO(3)$ are
\begin{eqnarray}
J_x=\left(
\begin{array}{ccc}
0 & 0 & 0 \\
0 & 0 & -i \\
0 & i & 0 \\
\end{array}
\right),&& J_y= \left(
\begin{array}{ccc}
0 & 0 & i \\
0 & 0 & 0 \\
-i & 0 & 0 \\
\end{array}
\right),\nonumber\\
 J_z&=&\left(
\begin{array}{ccc}
0 & -i & 0 \\
i & 0 & 0 \\
0 & 0 & 0 \\
\end{array}
\right).
\end{eqnarray} 
Let us adopt that there occurs a spontaneous symmetry breaking.
We choose the symmetry breaking term as
\begin{equation}
\mathcal{L}_{SB} = \lambda\phi_3.
\end{equation}
$\mathcal{L}_{SB}$ is invariant under the transformation
$\phi\rightarrow e^{iJ_z\theta_z}\phi$.  For the transformation
$\phi\rightarrow e^{iJ_x\theta_x}\phi$, however, we have
\begin{equation}
\delta\mathcal{L}_{SB} = -\lambda\theta_x\phi_2.
\end{equation}
Similarly,
\begin{equation}
\delta\mathcal{L}_{SB} = \lambda\theta_y\phi_1,
\end{equation}
under the transformation $\phi\rightarrow e^{iJ_y\theta_y}\phi$.
Then, we have two massless bosons $\phi_1$ and $\phi_2$ and
one massive scalar field $\phi_3$.
For example, This is easily seen for the potential
\begin{equation}
\mathcal{L} = \frac{1}{2}(\partial_{\mu}\phi_i)(\partial^{\mu}\phi_i)
-\frac{m^2}{2}\phi_i\phi_i-g(\phi_i\phi_i)^2,
\end{equation}
with $m^2<0$ and $g>0$
by expanding the potential $V$ around the minimum.
\\
\\
(c) Multi-component complex scalar field\\
A model with a complex multi-component scalar field exhibits
similar symmetry breaking.  Let us consider a complex
three-component scalar field theory given as
\begin{equation}
\mathcal{L} = \partial_{\mu}\phi^{\dag}\partial^{\mu}\phi
-\frac{m^2}{2}\phi^{\dag}\phi-g(\phi^{\dag}\phi)^2,
\end{equation}
where $\phi= ^t(\phi_1,\phi_2,\phi_3)$ and $g>0$.
The Lagrangian has a $SU(3)$ symmetry.
The bases are given by the Gell-Mann matrices $\lambda_a$
($a=1,\cdots,8$): $\{T_a=\lambda_a/2\}$\cite{geo99}.
We adopt $m^2<0$ and consider the symmetry breaking term given by
\begin{equation}
\mathcal{L}_{SB} = \lambda(\phi_3+\phi^{\dag}_3).
\end{equation}
This term is not invariant under the tranformation
$\phi\rightarrow e^{iT_a\theta_a}\phi$ for $a=4$, 5, 6, 7 and 8.
Thus, after spontaneous symmetry breaking, we have five massless
Nambu-Goldstone bosons and one massive scalar field.
The massive boson is $\phi_3+\phi^{\dag}_3$ with the mass
$8ga^2$ for $a=Re\langle\phi_3\rangle$, and massless bosons are
$\phi_1$, $\phi_2$ and $i(\phi_3-\phi^{\dag}_3)$. 
The number of massless bosons is obtained straightforwardly in
our formulation.

\subsubsection{Superconducting transition}

To discuss a superconducting transition, we use the Nambu representation
$\psi=^t(\psi_{\uparrow},\psi^{\dag}_{\downarrow})$.
Let us consider a non-relativistic model of superconductivity given by
\begin{equation}
\mathcal{L}= i\psi^{\dag}\partial_t\psi-\psi^{\dag}\sigma_3\xi(\nabla))\psi
-g\psi^{\dag}_{\uparrow}(x)\psi_{\uparrow}(x)\psi^{\dag}_{\downarrow}(x)
\psi_{\downarrow}(x),
\end{equation}
where the last term with the coupling constant $g<0$ is an attractive
interaction term.
This Lagrangian is invariant under the transformation
\begin{equation}
\psi\rightarrow e^{-i\theta\sigma_3}\psi.
\end{equation}
This is the $U(1)$ phase transformation: 
$\psi_{\sigma}\rightarrow e^{-i\theta}\psi_{\sigma}$.
We add the following symmetry breaking term
\begin{equation}
\mathcal{L}_{SB}= \lambda\psi^{\dag}\sigma_1\psi
= \lambda( \psi^{\dag}_{\uparrow}\psi^{\dag}_{\downarrow}
+\psi_{\downarrow}\psi_{\uparrow} ),
\end{equation}
so that the invariance is lost.
Then from the general theory the NG boson $\pi$ is given as
\begin{equation}
\pi= i\psi^{\dag}[\sigma_3, \sigma_1]\psi=
2i( \psi^{\dag}_{\uparrow}\psi^{\dag}_{\downarrow}
-\psi_{\downarrow}\psi_{\uparrow} ).
\end{equation}
$\pi$ is shown to be a massless boson following the argument 
in the subsection 2.4.

\subsubsection{A multi-component fermion model}

We can also consider a multi-component fermion field.
Let us consider a fermion triplet $^{t}\psi= (\psi_1,\psi_2,\psi_3)$.
The symmetry group is $G=SU(3)$ and $\{T_a\}$ are given by the
Gell-Mann matrices.
We add the symmetry breaking term
\begin{equation}
\mathcal{L}_{SB}= \lambda\psi^{\dag}T_3\psi,
\end{equation}
for the symmetry breaking $\langle\psi^{\dag}T_3\psi\rangle\neq 0$.
Then the system is invariant under the transformation by $T_3$ and $T_8$.
NG bosons are given by $\pi_a= i\psi^{\dag}[T_a, T_3]\psi$.
Because the structure constants are given by $f_{123}=1$ and
$f_{345}=-f_{367}=1/2$, there are three NG bosons:
\begin{equation}
\pi_1= \psi^{\dag}T_2\psi,~~ \pi_4= \frac{1}{2}\psi^{\dag}T_5\psi,~~
\psi_6= -\frac{1}{2}\psi^{\dag}T_7\psi.
\end{equation}
$\pi_2$, $\pi_5$ and $\pi_7$ are also NG bosons, but these are not
independent.

When the symmetry breaking is given by $\langle\psi^{\dag}T_8\psi\rangle\neq 0$,
the symmetry group $G$ reduces to $H=\{ T_1, T_2, T_3, T_8\}$.
In this case, we have two NG bosons.

\subsection{Proof that $\pi_a$ is an NG Boson}

The pole of the Green's function gives information on the
energy spectrum of the particle\cite{agd}.  Thus, we investigate the
Green's function in the following.

The normalization of $\{T_a\}$ is given as
\begin{equation}
{\rm Tr}T_aT_b = c\delta_{ab},
\end{equation}
where $c$ is a real constant: $c\in {\bf R}$.  The commutators are
\begin{equation}
[T_a, T_b]= \sum_c if_{abc}T_c,
\end{equation}
where $f_{abc}$ are structure constants of the Lie algebra $g$.
We use the relation
\begin{equation}
\sum_{ab}f_{abc}f_{abd}= C_2(G)\delta_{cd},
\end{equation}
where $C_2(G)$ indicates the Casimir invariant of the Lie group $G$.
$C_2(G)$ is given by
\begin{eqnarray}
C_2(G) &=& 2Nc ~~ {\rm for}~G=SU(N)\\
&=& (N-2)c ~~ {\rm for}~G=O(N).
\end{eqnarray}
For example, for $SU(2)$, we have $C_2(G)=8$ when we use $c=2$.
The above relation results in
\begin{equation}
\sum_a [T_a, [T_a, T_b]]=\sum_{acd}f_{acb}f_{acd}T_d= C_2(G)T_b.
\end{equation}

Let $M$ be an element of the basis set of $g$: $M=T_m \in \{T_a\}$.
From Eq.(\ref{qpsi}), we have
\begin{eqnarray}
&& e^{i\theta Q_a}\psi^{\dag}e^{-i\theta Q_a}[M, T_a]e^{i\theta Q_a}
\psi e^{-i\theta Q_a}\nonumber\\
&&~~ =\psi^{\dag}[M,T_a]\psi+i\theta\psi^{\dag}[T_a, [M, T_a]]\psi+O(\theta^2)
\nonumber\\
&&~~ =-i\sum_cf_{amc}\psi^{\dag}T_c\psi-i\theta \sum_{cd} f_{acm}f_{acd}
\psi^{\dag}T_d\psi +O(\theta^2). 
\nonumber\\
\end{eqnarray}
We assume that 
$\langle\psi^{\dag} T_m\psi\rangle=\langle\psi^{\dag} M\psi\rangle\neq 0$ and
$\langle\psi^{\dag} T_d\psi\rangle=0$ $(d\neq m)$.
Then, the order parameter is written as
\begin{eqnarray}
\Delta &=& \langle\psi^{\dag}M\psi\rangle \nonumber\\
&=& \lim_{\theta\to 0}\frac{i}{\theta \sum_cf_{acm}^2} \langle e^{i\theta Q_a}
\psi^{\dag} e^{-i\theta Q_a}[M, T_a] e^{i\theta Q_a}\psi
e^{-i\theta Q_a}\rangle \nonumber\\
&=& \lim_{\theta\to 0}\frac{i}{\theta \sum_cf_{acm}^2} \langle e^{i\theta Q_a}
\psi^{\dag} [M, T_a] \psi e^{-i\theta Q_a}\rangle \nonumber\\
&=& \lim_{\theta\to 0}\frac{-1}{\theta \sum_cf_{acm}^2} \langle e^{i\theta Q_a}
\pi_a e^{-i\theta Q_a} \rangle.
\end{eqnarray}
Here, because $Q_a$ is an operator (not matrix), we used
$e^{-i\theta Q_a}[M, T_a]e^{i\theta Q_a}= [M, T_a]$.

We write the Hamiltonian of the system as $H_0$ and add the symmetry
breaking term:
\begin{equation}
H_T \equiv H_0+H_{\lambda},
\end{equation}
where
\begin{equation}
H_{\lambda} = -\lambda \int d{\bf r}\psi^{\dag}M\psi.
\end{equation}
Let us denote the ground state of $H_T$ as $\phi$:
$H_T$: $H_T\phi=E\phi$.
From our assumption, $\phi$ is a unique ground state.
We consider
\begin{eqnarray}
A_a &\equiv& \langle\phi | e^{i\theta Q_a}\pi_a({\bf r}) e^{-i\theta Q_a}|
\phi\rangle \nonumber\\
&=& \langle \phi |e^{iH_Tt}e^{i\theta Q_a}\pi_a({\bf r}) e^{-i\theta Q_a}
e^{-iH_Tt}|\phi\rangle.
\end{eqnarray}
We use the notation $\tilde{\phi}=e^{-i\theta Q_a}\phi$ to write
\begin{equation}
A_a= \langle\tilde{\phi}| e^{-i\theta Q_a}e^{iH_Tt}e^{i\theta Q_a}
\pi_a({\bf r}) e^{-i\theta Q_a}e^{-iH_Tt}e^{i\theta Q_a}|\tilde{\phi}\rangle.
\end{equation}
Here we define the effective Hamiltonian $\tilde{H}$, using the
Campbell-Baker-Hausdorff formula:
\begin{eqnarray}
e^{-i\theta Q_a}e^{-iH_Tt}e^{i\theta Q_a}&=& \exp\left(-iH_Tt
+\theta t[H_T, Q_a] 
 +\cdots \right)\nonumber\\
&\equiv& \exp(-i\tilde{H}t).
\end{eqnarray}
Because of the relation
\begin{equation}
i[Q_a, \psi^{\dag}M\psi]= \pi_a,
\end{equation}
we obtain
\begin{eqnarray}
\tilde{H} &=& H_T+i\theta [H_T, Q_a]+O(\theta^2)\nonumber\\
&=& H_T+\theta\lambda \int d{\bf r}\pi_a({\bf r})
+O(\theta^2).
\end{eqnarray}
This results in
\begin{eqnarray}
A_a&=& \langle\tilde{\phi}| e^{i\tilde{H}t}\pi_a({\bf r}) e^{-i\tilde{H}t}|
\tilde{\phi}\rangle \nonumber\\
&=& \langle\tilde{\phi}| U(t,0)^{\dag}\pi_a({\bf r},t) U(t,0)|\tilde{\phi}
\rangle,
\end{eqnarray}
where $U(t,t')$ is given by
\begin{equation}
U(t,t')= e^{iH_Tt}e^{-i\tilde{H}(t-t')}e^{-iH_Tt'},
\end{equation}
and
\begin{equation}
\pi_a({\bf r},t)= e^{iH_Tt}\pi_a({\bf r})e^{-iH_Tt}.
\end{equation}
We can show that $\tilde{\phi}$ is an eigenstate of $\tilde{H}$:
$\tilde{H}\tilde{\phi}=E\tilde{\phi}$.
Hence, from the Gell-Mann-Low adiabatic theorem\cite{fetter}, we have
\begin{equation}
\tilde{\phi}= U(0,-\infty)\phi,
\end{equation}
where $\phi$ is the eigenstate of the Hamiltonian without the
perturbation by $\theta$ term, namely, the eigenstate of $H_T$.
This leads to
\begin{equation}
A_a= \langle\phi |U(t,-\infty)^{\dag}\pi_a({\bf r},t)U(t,-\infty)|
\phi\rangle.
\end{equation}
We defined a time-ordered exponential as
\begin{equation}
U(t,-\infty)= T\exp\left( -i\int_{-\infty}^t H_1(t')dt' \right),
\end{equation}
where
\begin{equation}
H_1 = -\theta\lambda \int d{\bf r}\pi_a({\bf r}),
\end{equation}
and we use the notation $H_1(t)=e^{iH_Tt}H_1e^{-iH_Tt}$.
$A_a$ is expanded in terms of $H_1$ as follows:
\begin{eqnarray}
A_a &=& \langle\phi|\pi_a |\phi\rangle
-i\int_{-\infty}^{t}dt' \langle\phi |[\pi_a({\bf r},t), H_1(t')]|
\phi\rangle +O(H_1^2) \nonumber\\
&=& \langle\phi|\pi_a |\phi\rangle
+i\theta\lambda \int d{\bf r}'\int_{-\infty}^t dt'
\langle\phi |[\pi_a({\bf r},t), \pi_a({\bf r}',t')]|\phi\rangle
\nonumber\\
&&~~ +O(H_1^2)\nonumber\\
&=& \langle\phi|\pi_a |\phi\rangle
-\theta\lambda\int d{\bf r}' \int_{-\infty}^{\infty}dt'
D_{aa}^R(t-t',{\bf r}-{\bf r}')\nonumber\\
&&~~ +O(H_1^2),
\end{eqnarray}
where we defined the retarded Green's function,
\begin{equation}
D_{aa}^R(t-t', {\bf r}-{\bf r}')= -i\theta(t-t')\langle 
[\pi_a({\bf r},t), \pi_a({\bf r}',t')]\rangle.
\end{equation}
By means of the Fourier transform given by
\begin{equation}
D_{aa}^R(\omega, {\bf r}-{\bf r}')= \int_{-\infty}^{\infty} dt
D_{aa}^R(t, {\bf r}-{\bf r}')e^{i\omega t},
\end{equation}
we obtain
\begin{equation}
A_a =\langle\phi|\pi_a |\phi\rangle
-\theta\lambda\int d{\bf r}'D^R_{aa}(\omega=0, {\bf r}-{\bf r}'),
\end{equation}
where the retarded Green's function $D_{aa}^R(\omega, {\bf r}-{\bf r}')$ 
is continued to the thermal Green's function
$D_{aa}(i\epsilon_n\rightarrow \omega+i\delta, {\bf r}-{\bf r}')$
taking the limit $\omega\rightarrow 0$, by analytic continuation\cite{agd}.
The thermal Green's function is given as
\begin{eqnarray}
D_{aa}(\tau-\tau', {\bf r}-{\bf r}')&=& -\langle T_{\tau}
\pi_a({\bf r},\tau)\pi_a({\bf r}',\tau')\rangle \nonumber\\
&=& \frac{1}{\beta}\sum_ne^{-i\epsilon_n(\tau-\tau')}
D_{aa}(i\epsilon_n, {\bf r}-{\bf r}').\nonumber\\
\end{eqnarray}
Here, $T_{\tau}$ is the time-ordering operator.
Because $\langle\phi|\pi_a |\phi\rangle=0$,
the gap function is written as
\begin{eqnarray}
\Delta &=& \lim_{\theta \to 0}\frac{-1}{\theta \sum_cf_{acm}^2} A_a\nonumber\\
&=& \frac{1}{\sum_cf_{acm}^2}\lambda  D_{aa}(\omega=0, {\bf q}=0),
\label{deltaD}
\end{eqnarray}
where $D_{aa}(\omega=0, {\bf q}=0)$ is the ${\bf q}=0$ and $\omega=0$
component of the Fourier transform of $D_{aa}(\omega, {\bf r}-{\bf r}')$.

When $\Delta$ is finite $(\neq 0)$ in the limit $\lambda\rightarrow 0$,
this formula indicates that $D_{aa}$ is given in the form
for small $\lambda$:
\begin{equation}
D_{aa}(\omega,{\bf q})=\frac{P_2(\omega,{\bf q})}
{a\lambda+P_1(\omega,{\bf q})},
\label{ngd}
\end{equation}
where $P_1$ and $P_2$ should satisfy $P_1(\omega,{\bf q})\rightarrow 0$
as ${\bf q}\rightarrow 0$ and $\omega\rightarrow 0$ and
$P_2(\omega,{\bf q})$ is a constant $\neq 0$ in the same limit.
$a$ is also a constant $(\neq 0)$ in this limit.
In the limit $\lambda\rightarrow 0$, $D_{aa}$ reads
$D_{aa}(\omega,{\bf q})=P_2(\omega,{\bf q})/P_1(\omega,{\bf q})$.
Since $P_1$ has a zero at $\omega=0$ and ${\bf q}=0$, we
have the dispersion relation $\omega({\bf q})$ satisfying
\begin{equation}
\omega({\bf q})\rightarrow 0~~ {\rm as}~~{\bf q}\rightarrow 0.
\end{equation}

For example, for a ferromagnet, we add the Zeeman term
to the Hamiltonian:
$H_{ferro}= -J\sum_{j\hat{\mu}}{\bf S}_j\cdot{\bf S}_{j+\hat{\mu}}
-H_z\sum_jS_{jz}$ where the vectors $\hat{\mu}$ connect the site
$j$ with its nearest neighbors on a lattice.
The term with $H_z$ breaks a rotational symmetry.
The dispersion relation for the spin wave excitation (NG mode) is
\begin{equation}
\omega({\bf q})= H_z+JS\sum_{\hat{\mu}}({\bf q}\cdot{\hat{\mu}})^2,
\end{equation}
for small $|{\bf q}\cdot{\hat{\mu}}|$.
This form is consistent with the form in Eq.(\ref{ngd}) when
we expand $P_1$ in terms of $\omega$ and ${\bf q}$.

In the normal phase where $\Delta$ vanishes as $\lambda\rightarrow 0$,
we have from Eq.(\ref{deltaD})
\begin{equation}
\chi_{\Delta}= D_{aa}(\omega=0,{\bf q}=0),
\label{chingd}
\end{equation}
where we scaled $\lambda$ so that the coefficient is unity.
It is sometime adopted that a fluctuation mode is written 
in the form
\begin{equation}
D^{-1}(\omega,{\bf q})\simeq \delta+P(\omega,{\bf q}),
\end{equation}
where $\delta$ indicates the distance from the transition
point and $P(\omega=0,{\bf q}=0)=0$.  
In the spin-fluctuation theory for a ferromagnet,
we use $P(\omega,{\bf q})=Aq^2+iC\omega/q$ at $T>T_c$
where $q=|{\bf q}|$, and $A$ and $C$ are constants\cite{mor85}. 
The Eq.(\ref{chingd}) results in
\begin{equation}
\delta= \chi_{\Delta}^{-1}.
\end{equation}

From Eq.(\ref{deltaD}), we obtain the expression,
\begin{equation}
\Delta= \frac{1}{C_2(G)}\lambda \sum_a' D_{aa}(\omega=0, {\bf q}=0),
\end{equation}
where $\sum_a'$ indicates that we do not include $T_a$ which commutes
with $M$.
The field $\pi_a=i\psi^{\dag}[T_a, M]\psi$ with 
$[M, T_a]\neq 0$ indicates the 
massless Nambu-Goldstone boson.
When there is the symmetry breaking term 
$\mathcal{L}_{SB}=\lambda\psi^{\dag}M\psi$, the symmetry is reduced from
$G$ to a subgroup $H$.  $[M, T_a]\neq 0$ means that $T_a$ is in
$G/H$.

\subsection{NG Boson Green's Functions and Vanishing Theorem}

Let us investigate the Nambu-Goldstone Green's functions given by
\begin{equation}
D_{ab}(x-y)= -i\langle T\pi_a(x)\pi_b(y)\rangle,
\end{equation}
for $x=(x_0=t, {\bf r})$.
Let $M=T_m$ and consider
\begin{eqnarray}
&& e^{i\theta Q_b}\psi^{\dag}e^{-i\theta Q_b}[T_m, T_a]
e^{i\theta Q_b}\psi e^{-i\theta Q_b} \nonumber\\ 
&&~~ = -i\sum_c f_{amc}\psi^{\dag}T_c\psi
+i\theta\sum_{cd}f_{amc}f_{bcd}\psi^{\dag}T_d\psi+O(\theta^2).
\nonumber\\
\end{eqnarray}
Because $\langle\psi^{\dag}T_c\psi\rangle=0$ $(c\neq m)$, we have
\begin{eqnarray}
&& \langle e^{i\theta Q_b}\psi^{\dag}e^{-i\theta Q_b}[T_m, T_a]
e^{i\theta Q_b}\psi e^{-i\theta Q_b}\rangle \nonumber\\
&& ~~~ = i\theta\sum_{cd}f_{amc}f_{bcd}\langle\psi^{\dag}T_d\psi\rangle
+O(\theta^2) \nonumber\\
&& ~~~ = -i\theta\sum_cf_{amc}f_{bmc}\langle\psi^{\dag}T_m\psi\rangle
+O(\theta^2).
\end{eqnarray}
We assume $a\neq b$ with $a\neq m$ and $b\neq m$.
There are two cases: (i) $\sum_cf_{amc}f_{bmc}\neq 0$ and
(ii) $\sum_cf_{amc}f_{bmc}=0$.

First, let us consider the case $\sum_cf_{amc}f_{bmc}\neq 0$.
Then we obtain
\begin{eqnarray}
\Delta &=& \langle\psi^{\dag} T_m\psi\rangle \nonumber\\
&=& \lim_{\theta\to 0}  \frac{-1}{\theta\sum_cf_{amc}f_{bmc}}A_{ab},
\end{eqnarray}
where
\begin{eqnarray}
A_{ab} &=& \langle\phi |\pi_a |\phi\rangle
-\theta\lambda \int d{\bf r}'\int_{-\infty}^{\infty} dt
D_{ab}^R (t-t',{\bf r}-{\bf r}') 
 +O(\theta^2) \nonumber\\
&=& \theta\lambda D_{ab}(\omega=0, {\bf q}=0)+O(\theta^2).
\end{eqnarray}
We put $\langle \phi| \pi_a|\phi\rangle=0$.
This leads to
\begin{equation}
\Delta= \frac{1}{\sum_cf_{amc}f_{bmc}}\lambda
D_{ab}(\omega=0, {\bf q}=0).
\end{equation}
This indicates that in the limit $\lambda\rightarrow 0$,
\begin{equation}
D_{ab}(\omega=0, {\bf q}=0)\propto \frac{1}{\lambda}.
\end{equation}
Hence the NG boson Green's function $D_{ab}(\omega, {\bf q})$ also has a pole
for $\omega\rightarrow 0$ and ${\bf q}\rightarrow 0$ if
$\sum_cf_{amc}f_{bmc}\neq 0$.

In real algebras, the condition $\sum_cf_{amc}f_{bmc}\neq 0$ sometimes
leads to that $\pi_a$ and $\pi_b$ are identical: $\pi_a=\pi_b$.
For example, let us consider $[T_a,T_m]= \beta T_c$,
$[T_b,T_m]=\gamma T_c$, $[T_a,T_b]=0$ and
$[T_c,T_m]=-\beta T_a-\gamma T_b$ for constants $\beta$ and $\gamma$.
In this case, $\pi_a\propto \psi^{\dag}T_c\psi$ is the same as
$\pi_b\propto \psi^{\dag}T_c\psi$, and we have two NG bosons $\pi_a$ and
$\pi_c=-i\beta\psi^{\dag}T_a\psi-i\gamma\psi^{\dag}T_b\psi$.

Now let us consider the case $\sum_cf_{amc}f_{bmc}=0$.
In this case we have
\begin{eqnarray}
&& \langle e^{i\theta Q_b}\psi^{\dag}e^{-i\theta Q_b}[T_m, T_a]
e^{i\theta Q_b}\psi e^{-i\theta Q_b}\rangle = 0 \nonumber\\
&& ~~ = i\langle e^{i\theta Q_b}\pi_a e^{-i\theta Q_b}\rangle.
\end{eqnarray}
This results in the vanishing property:
\begin{equation}
\int dt \int d{\bf r}D_{ab}(t, {\bf r})= D_{ab}(\omega=0, {\bf q}=0)= 0,
\end{equation}
if $\sum_cf_{amc}f_{bmc}=0$.
Thus we obtain the vanishing of the space-time integral of the NG boson Green's
function $D_{ab}(t, {\bf r})$ under the condition $\sum_cf_{amc}f_{bmc}=0$.
In this case $D_{ab}$ does not represent a massless mode.
 
The vanishing of the Green's function occurs, for example,  when three elements
of a basis set $\{ T_1, T_2, T_3\}$ are closed:
\begin{equation}
[T_a, T_b]= i\sum_c \epsilon_{abc}T_c,
\end{equation}
where $\epsilon_{abc}$ is the totally antisymmetric symbol with $\epsilon_{123}=1$.
We assume that $\langle\psi^{\dag}T_3\psi\rangle\neq 0$ and
$\langle\psi^{\dag}T_1\psi\rangle=\langle\psi^{\dag}T_2\psi\rangle=0$.
The NG bosons are given by $\pi_1=\psi^{\dag}T_2\psi$ and
$\pi_2= -\psi^{\dag}T_1\psi$.
Then, the propagators $D_{11}$ and $D_{22}$ represent massless modes and we have
\begin{equation}
\int dt\int d{\bf r}D_{12}(t, {\bf r})=0.
\end{equation}
This means that there is a constraint on $\pi_1$ and $\pi_2$ and that $\pi_1$
and $\pi_2$ are not independent.
Thus we have only one NG boson in this base.

\begin{figure}[htbp]
\begin{center}
  \includegraphics[height=2.2cm]{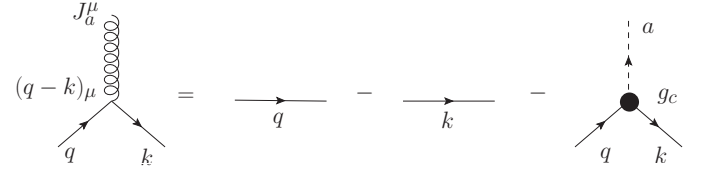}
\caption{Generalized Ward-Takahashi identity with the NG boson.
The straight line indicates the fermion propagator, and the coil-shaped line
shows the current $J_a^{\mu}$.  The dashed line indicates the NG boson
propagator.  The fermion-NG boson coupling (matrix) is denoted as $g_c$.
}
\label{fig1}       
\end{center}
\end{figure}

\section{Ward-Takahashi identity with NG Bosons}

\subsection{Modified Ward-Takahashi Identity}

We have conserved currents $J^{\mu}_a$ with $\partial_{\mu}J^{\mu}_a=0$
when $\mathcal{L}$ is invariant under some transformation.
When the symmetry is spontaneously broken, non-vanishing
$\partial_{\mu}J^{\mu}_a$ represents the Nambu-Goldstone boson.
For the transformation $\psi\rightarrow \psi-i\theta T_a\psi$, the
current is 
\begin{equation}
J_a^{\mu}= \psi^{\dag}\Gamma^{\mu}T_a\psi.
\end{equation}
Let us examine the expectation value
$\langle T(J^{\mu}_a(x)\psi(y)\psi^{\dag}(z))\rangle$ where
$x$ indicates the four vector $x=(x^0,{\bf r})$.
We evaluate the derivative of this expectation value:
\begin{eqnarray}
&&\partial_{\mu}^x \langle T(J^{\mu}_a(x)\psi(y)\psi^{\dag}(z))\rangle
 = \langle T(\partial_{\mu}^x J_a^{\mu}(x)\psi(y)\psi^{\dag}(z))\rangle
\nonumber\\
&&~~~ +\delta(x^0-y^0)\langle T([J_a^0(x),\psi(y)]\psi^{\dag}(z))\rangle \nonumber\\
&&~~~ +\delta(x^0-z^0)\langle T(\psi(y)[J_a^0(x),\psi^{\dag}(z)])\rangle.
\end{eqnarray}
We set $\Gamma^0= 1$ (unit matrix) for simplicity and 
we have 
\begin{equation}
J_a^0=\psi^{\dag}T_a\psi,~~~Q_a=\int d{\bf r}J_a^0.
\end{equation}
We use the commutation relations:
\begin{eqnarray}
\delta(x^0-y^0)[J_a^0(x),\psi(y)]&=& -\delta(x-y)T_a \psi(x)\\
\delta(x^0-y^0)[J_a^0(x),\psi^{\dag}(y)]&=& \delta(x-y)\psi^{\dag}(x)T_a.
\end{eqnarray}
This results in the following equation,
\begin{eqnarray}
&&\partial_{\mu}^x \langle T(J^{\mu}_a(x)\psi(y)\psi^{\dag}(z))\rangle
 = \lambda\langle T(\pi_a(x)\psi(y)\psi^{\dag}(z))\rangle \nonumber\\
&&~~~ -\delta(x-y)T_a\langle T(\psi(x)\psi^{\dag}(z))\rangle 
\nonumber\\
&&+~~~  +\delta(x-z)\langle T(\psi(y)\psi^{\dag}(z))\rangle T_a .
\end{eqnarray}
This is the Ward-Takahashi identity with the Nambu-Goldstone boson.

We define the Fourier transforms of correlation functions.
We introduce the vertex function $\Gamma_a^{\mu}$:
\begin{eqnarray}
&& \int d^4x\int d^4y\int d^4z e^{-ip\cdot x-ik\cdot y+iq\cdot z}
\partial_{\mu}^x \langle TJ_a^{\mu}(x)\psi(y)\psi^{\dag}(z)\rangle
\nonumber\\
&&~~~ \equiv ip_{\mu}i(2\pi)^4\delta^4(p+k-q) G(k)\Gamma_a^{\mu}(k,q) G(q),
\end{eqnarray}
where $G(k)$ is the Green's function of fermion $\psi$ given by
\begin{equation}
i(2\pi)^4\delta^4(k-q)G(k) \equiv 
 \int d^4y \int d^4z
e^{-ik\cdot y+iq\cdot z} \langle T(\psi(y)\psi^{\dag}(z))\rangle.
\end{equation}
$k$ is the four momentum $k=(k^0,{\bf k})$.
There appears the expectation value 
$\langle T(\pi_a(x)\psi(y)\psi^{\dag}(z))\rangle$ that contains $\pi_a$.
The interaction between fermions would induce an effective interaction
between $\pi_a$ and fermions.  Thus we introduce the fermion-NG boson
coupling (vertex function) $g_a(k,q)$:
\begin{eqnarray}
&& \int d^4x \int d^4y \int d^4z e^{-ip\cdot x-ik\cdot y+iq\cdot z}
\langle T(\pi_a(x)\psi(y)\psi^{\dag}(z))\rangle\nonumber\\
&&~~~ = (2\pi)^4\delta^4(p+k-q)\big[ \sum_bf_{amb}G(k)T_bG(q)\nonumber\\
&&~~~ +\sum_cG(k)g_c(k,q)G(q)D_{ca}(q-k) \big],
\label{vertex}
\end{eqnarray}
where the summation with respect to $c$ is taken for which $D_{ca}$
does not vanish.
Because $\pi_a$ is in general a linear combination of $\psi^{\dag}T_a\psi$, we can
consider $\langle T(\psi^{\dag}(x)T_a\psi(x)\psi(y)\psi^{\dag}(z))\rangle$.
We define 
\begin{equation}
B_{ab}(x-y)= -i\langle T(\psi^{\dag}T_a\psi)(x)(\psi^{\dag}T_b\psi)(y)\rangle,
\end{equation}
and its Fourier transform given as
\begin{equation}
B_{ab}(k)= \int d^4x e^{-ik\cdot (x-y)} B_{ab}(x-y).
\end{equation}
In the non-interacting case, $B_{ab}(k)$ is
\begin{equation}
B_{ab}^{0}(q)= -i\int\frac{d^4k}{(2\pi)^4}{\rm Tr}T_aG^0(k)T_bG^0(k+q).
\end{equation}
Because we have
\begin{equation}
D_{ab}(q)= \sum_{cd}f_{acm}f_{bdm}B_{cd}(q),
\end{equation}
we set
\begin{equation}
f_c(k,q)= \sum_a f_{amc}g_a(k,q),
\end{equation}
to obtain
\begin{eqnarray}
&& \int d^4x d^4y d^4z e^{-ip\cdot x-ik\cdot y+iq\cdot z}
\langle T(\psi^{\dag}(x)T_a\psi(x)
 \psi(y)\psi^{\dag}(z))\rangle\nonumber\\
&&~~ = (2\pi)^4\delta^4(p+k-q)\big[ -G(k)T_aG(q) \nonumber\\ 
&&~~~  -\sum_cG(k)f_c(k,q)B_{ca}(q-k)G(q) \big].
\nonumber\\
\end{eqnarray}

In the momentum space, the Ward-Takahashi identity is written in 
the form:
\begin{eqnarray}
(q-k)_{\mu}G(k)\Gamma_a^{\mu}(k,q)G(q)
 &=& iT_aG(q)-iG(k)T_a \nonumber\\
&& -\lambda G(k)\Gamma_a(k,q)G(q),
\nonumber\\
\end{eqnarray}
where
\begin{equation}
\Gamma_a(k,q)= \sum_cf_{amc}T_c
+\sum_cg_c(k,q)D_{ca}(q-k).
\end{equation}
This is diagrammatically shown in Fig.1 and is written as
\begin{equation}
(q-k)_{\mu}\Gamma_a^{\mu}(k,q)
 =iG^{-1}(k)T_a -iT_aG^{-1}(q) -\lambda \Gamma_a(k,q).
\end{equation}
This is the modified Ward-Takahashi identity with the correction
from the Nambu-Goldstone boson.  
We have let that $M=T_m\in \{T_c\}$.
Because $\lambda D_{ab}(q-k\rightarrow 0)= \sum_cf_{amc}f_{bmc} \Delta$ as
$\lambda\rightarrow 0$,
we obtain 
\begin{equation}
\lambda\Gamma_a(k,k)\rightarrow \sum_{cd}g_c(k,k)f_{cmd}f_{amd}\Delta,
\end{equation}
as $\lambda\rightarrow 0$.
Then we have the relation
\begin{equation}
iG^{-1}(k)T_a-iT_aG^{-1}(k)-\Delta \sum_cg_c(k,k)\alpha_{ca}=0,
\label{wtd}
\end{equation}
with $\alpha_{ca}= \sum_df_{amd}f_{cmd}$.
The Green's function $G(k)$ is expressed as
\begin{equation}
G^{-1}(k)= k_0-{\bf \Gamma}\cdot {\bf k}-\Sigma(k),
\end{equation}
where we put $\Gamma^{\mu}=(\Gamma^0=1,{\bf \Gamma})$, and the above
relation results in
\begin{equation}
iT_a\Sigma(k)-i\Sigma(k)T_a+\Delta \sum_cg_c(k,k)\alpha_{ca}=0.
\end{equation}
When the interaction term is explicitly given, the self-energy
$\Sigma(k)$ and the vertex $\Gamma_a$ can be calculated.
This relation gives the equation for the order parameter $\Delta$ and
the coupling constant $g_c$.

\begin{figure}[htbp]
\begin{center}
  \includegraphics[height=2.2cm]{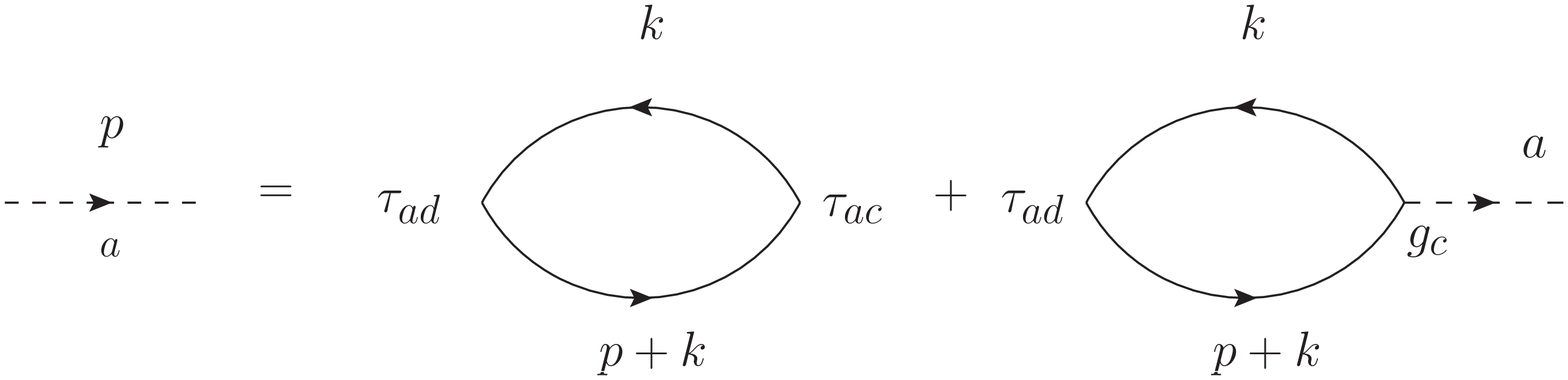}
\caption{Equation for the NG boson Green's function.
The solid line indicates the fermion propagator, and the dashed line
shows the NG boson propagator. 
$\tau_{ad}$ indicates $\tau_{ad}=\sum_{d}f_{amd}T_d$.
}
\label{fig2}       
\end{center}
\end{figure}

\subsection{Vertex Function for NG boson Green's Functions}

Let us investigate the equations for Nambu-Goldstone Green's functions.
First note that $\pi_a=i\psi^{\dag}[T_a,T_m]\psi=-\sum_cf_{amc}\psi^{\dag}T_c\psi$.
From Eq.(\ref{vertex}), we have
\begin{eqnarray}
&&\langle T \pi_a(x)\psi(y)\psi^{\dag}(z)\rangle \nonumber\\
&=& \int \frac{d^4p}{(2\pi)^4}\frac{d^4k}{(2\pi)^4}
e^{ip\cdot x+ik\cdot y-i(p+k)\cdot z} \nonumber\\
&& \times \Big[
\sum_cf_{amc}G(k)T_c G(p+k) \nonumber\\
&&+\sum_cG(k)g_c(k,k+p)D_{ca}(p)G(k+p) \Big].
\nonumber\\
\end{eqnarray}
This indicates
\begin{eqnarray}
&&{\rm Tr}\sum_cf_{amc}T_c\langle T\pi_a(x)\psi(y)\psi^{\dag}(y_0+\delta,
{\bf y})\rangle \nonumber\\
&=& -\sum_cf_{amc}\langle T\pi_a(x)\psi^{\dag}(y_0+\delta,{\bf y})
T_c\psi(y)\rangle \nonumber\\
&=& \langle T\pi_a(x)\pi_a(y)\rangle,
\end{eqnarray}
where we use 
${\rm Tr}T_c\pi_a(x)\psi(y)\psi^{\dag}(y)={\rm Tr}\pi_a(x)T_c\psi(y)
\psi^{\dag}(y)=-\pi_a(x)\psi^{\dag}(y)T_c\psi(y)$ because $\pi_a(x)$
is an operator (not a matrix).
Then the NG boson Green's function is given by
\begin{eqnarray}
&& \langle T\pi_a(x)\pi_a(y)\rangle \nonumber\\
&=& \int\frac{d^4p}{(2\pi)^4} \int\frac{d^4k}{(2\pi)^4}
e^{ip\cdot(x-y)}{\rm Tr}\sum_d f_{amd}T_d \nonumber\\
&& \times \Big[
 \sum_cf_{amc}G(k)T_cG(p+k) \nonumber\\
&& +\sum_cG(k)g_c(k,k+p) 
 D_{ca}(p)G(k+p) \Big].
\end{eqnarray}
This reads
\begin{eqnarray}
D_{aa}(p)&=& -i{\rm Tr}\sum_{cd}\int\frac{d^4k}{(2\pi)^4}\Big[
f_{amd}f_{amc}T_dG(k)T_cG(k+p) \nonumber\\
&+& f_{amd}T_dG(k)g_c(k,k+p)G(k+p)D_{ca}(p) \Big].
\end{eqnarray}
This is shown diagrammatically in Fig.2.
The Green's function for different NG bosons $\pi_a$ and $\pi_b$ is
\begin{eqnarray}
D_{ab}(p)&=& -i{\rm Tr}\sum_{cd}\int\frac{d^4k}{(2\pi)^4}\Big[
f_{bmd}f_{amc}T_dG(k)T_cG(k+p) \nonumber\\
&+& f_{bmd}T_dG(k)g_c(k,k+p)G(k+p)D_{ca}(p) \Big].
\end{eqnarray}
When $\sum_cf_{amc}f_{bmc}=0$ for $a\neq b$, we neglect $D_{ab}$ ($a\neq b$)
because $D_{ab}(p)\rightarrow 0$ as $p\rightarrow 0$.
In this case, the equation for $D_{aa}(p)$ reads
\begin{eqnarray}
D_{aa}(p)
&=& \Big[ 1+i{\rm Tr}\int\frac{d^4k}{(2\pi)^4}\sum_df_{amd}T_dG(k)
\nonumber\\
&&\times g_a(k,k+p)G(k+p)\Big]^{-1}\nonumber\\
&\times& (-i){\rm Tr}\sum_{cd}\int\frac{d^4k}{(2\pi)^4}f_{amd}f_{amc}T_d
G(k)T_cG(k+p).
\nonumber\\
\end{eqnarray}
$g_a(k.k+p)$ should be determined on the basis of the Ward-Takahashi
identity.

\subsection{Higgs boson}

We define the Higgs field $h(x)$ by
\begin{equation}
h(x)= \psi^{\dag}(x)T_m\psi(x),
\end{equation}
where $T_m$ is the basis corresponding to broken symmetry.
The Higgs boson indicates the fluctuation of the amplitude of
the order parameter $\Delta=\langle\psi^{\dag}T_m\psi\rangle$.
Thus, in a strict sense, the Higgs field should be defined as
\begin{equation}
\delta h(x) = \psi^{\dag}(x)T_m\psi(x)-\Delta.
\end{equation}
We simply call the field $h(x)$ the Higgs field.
$h(x)$ is composed of fermions as in the case of NG bosons.
Thus the Green's function of the Higgs boson,
\begin{eqnarray}
H(x-y)&=& -i\langle T(h(x)h(y))\rangle\nonumber\\
&=& \int\frac{d^4p}{(2\pi)^4}e^{ip\cdot(x-y)}H(p),
\end{eqnarray}
is also evaluated in a similar way to that of Nambu-Goldstone bosons.
We introduce the vertex function $g_H(k,k+p)$ to write
\begin{eqnarray}
H(p)&=& -i{\rm Tr}\int\frac{d^4k}{(2\pi)^4}\Big[ T_mG(k)T_mG(k+p)
\nonumber\\
&+& T_mG(k)g_H(k,k+p)G(k+p)H(p) \Big].
\label{higgsg}
\end{eqnarray} 
The vertex function $g_H(k,k+p)$ will depend on the interaction
between electrons.
It is reasonable to assume that $g_H(k,k+p)$ is proportional to
$T_m$ since $h(x)=\psi^{\dag}T_m\psi$.
Thus we denote $g_H(k,k+p)=g_m(k,k+p)$.
The dispersion of the Higgs boson is determined by this
equation.

\subsection{NG Boson-NG boson and NG Boson-Higgs Boson Couplings}

Because we have the NG boson-fermion coupling and the Higgs-fermion coupling,
there are NG boson-NG boson coupling and NG boson-Higgs coupling as
effective interactions.
The figures 3(a) and 3(b) indicate couplings of two and three particles,
respectively.  Multi-particle couplings also possibly exist.
When the Lagrangian including the interaction term is given, we can 
evaluate multi-particle vertex functions using some calculation
methods.

The figure 3(a) shows NG boson-NG boson coupling or NG boson-Higgs boson
coupling.  
In general, the NG boson-Higgs boson coupling $\pi_a h$ vanishes
because of the orthogonality of bases $T_a$: ${\rm Tr}T_aT_b=c\delta_{ab}$.

\begin{figure}[htbp]
\begin{center}
  \includegraphics[height=2.8cm]{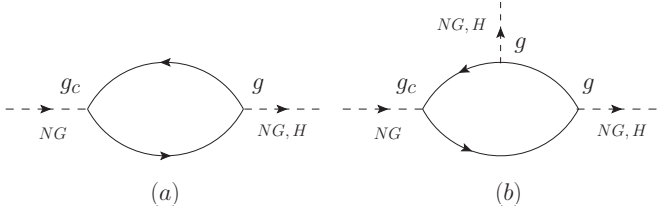}
\caption{NG boson-NG boson (or Higgs boson) couplings.
The solid line indicates the fermion propagator, and the dashed line
shows the NG boson or the Higgs boson propagator. 
(b) shows the coupling such as $\pi_a\pi_b h$.
$g_{c}$ and $g$ indicate the vertex functions where $g=g_d$ or
$g=g_H$.
}
\label{fig3}       
\end{center}
\end{figure}

\subsection{Some Physical Systems}

\subsubsection{Ferromagnetic transition}

We take $G=SU(2)$ and a fermion doublet $\psi=^t(\psi_{\uparrow},\psi_{\downarrow})$.
Let us consider the Hubbard model\cite{hub63,yam98,yan01,yan16b}
\begin{equation}
\mathcal{L}= i\psi^{\dag}\partial_t\psi-\psi^{\dag}\xi(\nabla)\psi
-U\psi^{\dag}_{\uparrow}(x)\psi_{\uparrow}(x)\psi^{\dag}_{\downarrow}(x)
\psi_{\downarrow}(x),
\end{equation}
where $\xi(\nabla)=\epsilon(\nabla)-\mu$ is the electron dispersion
relation with chemical potential $\mu$ and last term indicates the repulsive
interaction ($U>0$).
The bases $\{T_a\}$ are given by Pauli matrices: $T_a=\sigma_a$ ($a=$ 1, 2 and 3).
The structure constants are $f_{abc}=2\epsilon_{abc}$.
This Lagrangian is invariant under the transformations
\begin{equation}
\psi\rightarrow e^{-i\theta\sigma_a}\psi,~~~ (a=1, 2, 3).
\end{equation}
The symmetry breaking term is given by the magnetization of electrons for
a ferromagnetic transition:
\begin{equation}
\mathcal{L}_{SB}= \lambda \psi^{\dag}\sigma_3\psi= \lambda 
(\psi^{\dag}_{\uparrow}\psi_{\uparrow}-\psi^{\dag}_{\downarrow}\psi_{\downarrow}).
\end{equation}
This term breaks the symmetry $\psi\rightarrow e^{-i\theta\sigma_a}\psi$ for
$a=1$, 2.
The corresponding Nambu-Goldstone bosons are
\begin{eqnarray}
\pi_1&=& i\psi^{\dag}[\sigma_1,\sigma_3]\psi= 2\psi^{\dag}\sigma_2\psi,\nonumber\\
\pi_2&=& i\psi^{\dag}[\sigma_2,\sigma_3]\psi=-2\psi^{\dag}\sigma_1\psi.
\end{eqnarray}
The excitation mode represented by $\pi_1$ and $\pi_2$ is spin-flip process,
that is, the spin-wave excitation.
We make a linear combination of $\pi_1$ and $\pi_2$ as
$\pi \equiv (i\pi_1-\pi_2)/4= \psi^{\dag}_{\uparrow}\psi_{\downarrow}$ and 
$\pi^{\dag} = (-i\pi_1-\pi_2)/4= \psi^{\dag}_{\downarrow}\psi_{\uparrow}$. 
Actually, there is only one Nambu-Goldstone boson $\pi$ in a ferromagnetic state.
This is consistent with the general theory for counting the number of
NG bosons\cite{wat12,hid13} and also with the vanishing theorem.
As shown in the section II, $\pi$ represents a massless excitation.  
 
The electron Green's function is given in the form:
\begin{eqnarray}
G= \left(
\begin{array}{cc}
G_{\uparrow\uparrow} & 0 \\
0 & G_{\downarrow\downarrow} \\
\end{array}
\right),
\end{eqnarray}
where 
\begin{equation}
G_{\sigma\sigma}(x-y)= -i\langle T\psi_{\sigma}(x)\psi^{\dag}_{\sigma}(y)
\rangle.
\end{equation}
The self-energy $\Sigma$ is similarly defined as
\begin{eqnarray}
\Sigma= \left(
\begin{array}{cc}
\Sigma_{\uparrow} & 0 \\
0 & \Sigma_{\downarrow} \\
\end{array}
\right),
\end{eqnarray}
where $G_{\sigma\sigma}^{-1}(k)=k_0-\xi({\bf k})-\Sigma_{\sigma}$.
From the Ward-Takahashi identity in Eq.(\ref{wtd}), we obtain
\begin{eqnarray}
iG^{-1}\sigma_1-i\sigma_1G^{-1}-\Delta g_1f_{123}^2 &=& 0,\\
iG^{-1}\sigma_2-i\sigma_2G^{-1}-\Delta g_2f_{123}^2 &=& 0.
\end{eqnarray}
We set $g_a= \sum_c\epsilon_{acm}\sigma_c \tilde{g}$ such as 
$g_1= \sigma_2\tilde{g}$ and $g_2=-\sigma_1\tilde{g}$.
Making a linear combination $\sigma_2-i\sigma_1$, the above relation
results in
\begin{equation}
\Sigma_{\downarrow}(k)-\Sigma_{\uparrow}(k)= 
\Delta f_{123}^2 \tilde{g}(k,k).
\label{self-g}
\end{equation}
This is the relation between the electron-NG boson coupling and 
the self-energy.
When the self-energy is evaluated, the coupling constant $\tilde{g}$
is determined from this relation. 
This relation can be also regarded as the gap equation for $\Delta$.

Because $\pi= i(\pi_1+i\pi_2)/4$, the correlation function in
Eq.(\ref{vertex}) leads to
\begin{eqnarray}
&& \int d^4x \int d^4y \int d^4z e^{-ip\cdot x-ik\cdot y+iq\cdot z}
\langle T(\pi(x)\psi(y)\psi^{\dag}(z))\rangle\nonumber\\
&&~~~ = (2\pi)^4\delta^4(p+k-q)\Big[ -G(k)
\left(
\begin{array}{cc}
0 & 1 \\
0 & 0 \\
\end{array}
\right)
G(q)\nonumber\\
&&~~~ -4G(k)
\left(
\begin{array}{cc}
0 & 1 \\
0 & 0 \\
\end{array}
\right)G(q)\tilde{g}(k,q)\tilde{D}(q-k) \Big],
\end{eqnarray}
where $\tilde{D}(k)$ is the Fourier transform of the Green's function
of $\pi$:
\begin{equation}
\tilde{D}(x-y)= -i\langle T\pi(x)\pi^{\dag}(y)\rangle.
\end{equation}
Here we used the relation $\tilde{D}=D_{11}/8=D_{22}/8$.
When we calculate the Green's function 
$\langle T(\pi(x)\psi(y)\psi^{\dag}(z))\rangle$ by means of the perturbation
in Coulomb interaction $U$, the correction of the order of $U$ is 
\begin{eqnarray}
-G(k)
\left(
\begin{array}{cc}
0 & 1 \\
0 & 0 \\
\end{array}
\right)
G(q)-UG(k)
\left(
\begin{array}{cc}
0 & 1 \\
0 & 0 \\
\end{array}
\right)G(q)\tilde{D}(q-k), \nonumber\\
\end{eqnarray}
in the momentum space.  This gives
\begin{equation}
\tilde{g}(k,q)= \frac{U}{4}+\cdots.
\end{equation}
This is consistent with the self-energy-coupling relation in
Eq.(\ref{self-g}) since the self-energy is given by
$\Sigma_{\sigma}= Un_{-\sigma}+\cdots$ where $n_{\sigma}$ is the density
of electrons with spin $\sigma$, and we have 
$\Delta=n_{\uparrow}-n_{\downarrow}$.

\subsubsection{Superconductivity}

We obtain the Ward-Takahashi identity for 
superconductors in a similar way\cite{koy14,koy16}. 
The Higgs field $h$ is defined as
\begin{equation}
h= \psi^{\dag}\sigma_1\psi = \psi^{\dag}_{\uparrow}\psi^{\dag}_{\downarrow}
+\psi_{\downarrow}\psi_{\uparrow}.
\end{equation}
Near the critical temperature, the effective action for $h$ is given by
the time-dependent Ginzburg-Landau (TDGL) action with the dissipation
effect.
The Higgs mode in a superconductor is clearly defined at low temperatures
($T\ll T_c$).
The Higgs Green's function is given by
\begin{equation}
P_{11}(\omega,{\bf q})\equiv -i\frac{1}{2}{\rm Tr}\sigma_1G_0(\epsilon,{\bf k})
\sigma_1G_0(\epsilon+\omega,{\bf k}+{\bf q}),
\end{equation}
where $G_0$ is the electron Green's function:
\begin{eqnarray}
G_0^{-1}(\epsilon,{\bf k})=\left(
\begin{array}{cc}
\epsilon-\xi({\bf k}) & -\Delta \\
-\Delta & \epsilon+\xi({\bf k}) \\
\end{array}
\right),
\end{eqnarray}
where $\Delta$ is assumed to be real.
$1/g+P_{11}(\omega,{\bf q}=0)$ has a zero at $\omega=2\Delta$\cite{koy16}. 
At absolute zero, for small $\omega$ and $q=|{\bf q}|$, we obtain
\begin{equation}
\frac{1}{g}+P_{11}(\omega,{\bf q})= N(0)\Big[ 1-\frac{1}{3}
\left(\frac{\omega}{2\Delta}\right)^2\Big]+N(0)\frac{1}{3}c_s^2
\left(\frac{q}{2\Delta}\right)^2,
\end{equation}
where we adopt the approximation that the density of states is constant
and we used the gap equation,
\begin{equation}
\frac{1}{g} = N(0)\int d\xi\frac{1}{2E(\xi)}.
\end{equation}
We put $c_s^2=v_F^2/3$.

The relativistic model of superconductivity is given by the
Nambu-Jona-Lasinio model\cite{nam61}:
\begin{equation}
\mathcal{L}= \bar{\psi}i\gamma^{\mu}\partial_{\mu}\psi
+g_{NJL}[ (\bar{\psi}\psi)^2-(\bar{\psi}\gamma_5\psi)^2].
\end{equation}
This Lagrangian is invariant under the particle number and
Chiral transformations:
\begin{eqnarray}
&& \psi\rightarrow e^{i\theta}\psi,~~ \bar{\psi}\rightarrow 
\bar{\psi}e^{-i\theta} \\
&& \psi\rightarrow e^{i\gamma_5\theta}\psi,~~
\bar{\psi}\rightarrow \bar{\psi}e^{i\gamma_5\theta}.
\end{eqnarray}
The symmetry breaking term is
\begin{equation}
\mathcal{L}_{SB}= \lambda \bar{\psi}\psi,
\end{equation}
with $M=\gamma_0$.  Then the invariance under the transformation
$\psi\rightarrow \exp(i\gamma_5\theta)\psi$ is violated, and 
it is clear from our general theory that the NG boson and Higgs 
boson are given by
\begin{equation}
\pi= i\bar{\psi}\gamma_5\psi,~~~ h=\bar{\psi}\psi.
\end{equation}

\section{Summary}

We have given a formulation of the Nambu-Goldstone boson in fermion and
boson systems with spontaneous symmetry breaking.
The Nambu-Goldstone bosons are determined when the order
parameter in the phase transition is given in a system with
a continuous symmetry.
The Nambu-Goldstone boson $\pi_a$ is explicitly given by the formula
$\pi_a=i\psi^{\dag}[T_a, T_m]\psi$ for a fermion field $\psi$ where 
$T_a$ and $T_m$ are elements 
of basis set of the Lie algebra, where $T_m$ corresponds to the broken
symmetry.
We have given a proof that $\pi_a$ is a boson with vanishing mass
by showing that the susceptibility $\chi_{\Delta}$ is proportional 
to the NG boson
Green's function at $\omega=0$ and ${\bf q}=0$:
$\chi_{\Delta}\propto D_{aa}(\omega=0,{\bf q}=0)$.

When $\sum_cf_{amc}f_{bmc}=0$ holds, the vanishing property holds
where the Green's function $D_{ab}(q)$ of $\pi_a$ and $\pi_b$, given by
the Fourier transform of $\langle T\pi_a(x)\pi_b(y)\rangle$,
vanishes in the limit $q\rightarrow 0$: $D_{ab}(q=0)=0$.
This means that two bosons $\pi_a$ and $\pi_b$ are not independent
and there is a constraint.

The Ward-Takahashi identity is generalized in the presence of
spontaneous symmetry breaking.  The violation of the conservation
of the current is compensated by the inclusion of a contribution
from the Nambu-Goldstone boson.
We introduced the NG boson-fermion vertex function in the
Ward-Takahashi identity.
With this vertex function, the equation for NG boson Green's functions
is closed.
The NG boson-NG boson couplings and NG boson-Higgs boson couplings
are also introduced due to the NG boson-fermion and
Higgs boson-fermion vertex functions.

The Nambu-Goldstone boson degrees of freedom lead to the effective 
Lagrangian.  They describe the spin wave in magnetic 
systems\cite{kit87,tsv03} and the effective model is in general given
by the non-linear sigma model\cite{tsv03,leu94,leu94b}.
In superconductors, the effective action is given by the
sine-Gordon model\cite{koy96,yan12,yan13,nit15,yan16}.
We expect that
the coupling between NG bosons and fermions can be determined
on the basis of the Ward-Takahashi identity.

\section*{Acknowledgments}
The author thanks K. Odagiri for valuable discussions.
This work was supported by Grant-in-Aid for Scientific Research from the
Ministry of Education, Culture, Sports, Science and Technology in Japan
(Grant No. 17K05559).

\end{document}